\documentclass[twocolumn,twoside,slac_two]{revtex4}
\usepackage{graphicx}
\usepackage{fancyhdr}
\newcommand{\diff}{{\rm d}}
\newcommand{\lesim}{\,\raisebox{-0.4ex}{$\stackrel{<}{\scriptstyle\sim}$}\,}
\newcommand{\gesim}{\,\raisebox{-0.4ex}{$\stackrel{>}{\scriptstyle\sim}$}\,}
\newcommand{\eqb}{\begin{eqnarray}}
\newcommand{\eqe}{\end{eqnarray}}

\newcommand{\apj}{ApJ}
\newcommand{\araa}{Ann.\ Rev.\ Astron.\ Astrophys.}
\newcommand{\jgr}{J.\ Geophys.\ Res.}

\newcommand{\apjl}{ApJL}
\newcommand{\aap}{A\&A}
\newcommand{\mnras}{MNRAS}
\newcommand{\prd}{Phys.\ Rev.\ D}

\begin{document}

\title{Particle Acceleration in Relativistic Flows}

%

\author{J. G. Kirk}
\affiliation{Max-Planck-Institut f\"ur Kernphysik, 
69029 Heidelberg, Germany}

\begin{abstract}
A property common to several different astrophysical sources of high-energy
gamma-rays is the presence of bulk motion at relativistic speed. The intrinsic
spectra of the nonthermal radiating particles also show interesting
similarities, with a pronounced hardening towards lower energies.
This suggests two distinct
acceleration mechanisms could be at work in these sources. At high energies, 
the stochastic first-order Fermi process at shocks seems to provide a
reasonable explanation. I will briefly review the status of this mechanism
before discussing the possibility that, at lower energies, non-stochastic 
acceleration in the induced electric field of a relativistic current sheet
plays a role.
\end{abstract}

\maketitle

\thispagestyle{fancy}


\section{INTRODUCTION}
With only one or two exceptions, the identification of relativistic bulk
motion in an astrophysical source rests on the interpretation of its
nonthermal radiation  --- normally a featureless continuum, presumably 
produced by relativistic electrons as either 
synchrotron radiation or by the inverse Compton scattering mechanism. 
This makes it very difficult to constrain the underlying physics. In most
cases, cut-off or break frequencies are only poorly defined by the data
and the only quantity that can be used to constrain models is a 
spectral index. Even this is not always easy to interpret: to be meaningful, 
it requires a power-law spectrum extending over at least
a couple of decades in frequency. 
Nevertheless, the evidence accumulated from the spectra of shocked pulsar
winds, gamma-ray bursts,
and blazars indicates that the distribution of 
radiating particles is relatively soft at high energy 
($s=-\diff\ln f/\diff\ln \gamma\gesim4$, where $f$ is the phase-space density
of the injected particles) and
hard at low frequencies ($s\lesim3.6$).
The first-order Fermi mechanism operating at a relativistic shock 
predicts indices not too different from those seen at high energies, but it
fails to provide an explanation for the hard low energy spectra. A promising
candidate mechanism in this regime appears to be acceleration at current
sheets. In this paper I briefly discuss the observational evidence, review the
predictions of the first-order Fermi mechanism and summarise recent ideas on
relativistic current sheets. 

\section{OBSERVATIONS}
Plerions --- filled-centre supernova remnants \citep{weilerpanagia78} 
--- are thought to contain a
pulsar that powers the nonthermal nebular emission within the supernova 
bow-shock via a magnetised, relativistic wind.
Although data on the spectra of several examples have been modelled
\citep[e.g.,][]{chevalier00}, by far the best observed 
member of this class is the Crab Nebula. 
Optical and X-ray images of this object 
show features that apparently move with 
mildly relativistic velocities $\sim0.5 c$ \citep{hesteretal02}.
However, an analysis of the dynamics of the pulsar wind responsible for the
powering of the Nebula indicates that it is highly relativistic (bulk 
Lorentz factor $\Gamma\sim10^3$ to $10^6$) 
and may also be strongly magnetised
\citep{reesgunn74,kundtkrotscheck80,kennelcoroniti84b,lyubarskykirk01,lyubarsky03b}. Over the
entire X-ray energy range $2\times10^{16}$ to $6\times10^{20}\,$Hz 
its spectrum is
well-described by a power-law of photon index 
$q=-\diff \ln N_\gamma/\diff\ln\diff\nu=2.11$
\citep{massaroetal00}. In the gamma-ray region, this synchrotron spectrum cuts
off at around $10^{22}\,$Hz, whereas towards lower frequencies it hardens
substantially, reaching a photon index of $q=0.26$ in the radio range 
($10^8$ to $10^{10}\,$Hz),
\citep{baarshartsuijker72,bietenholzkronberg92}.

In the case of gamma-ray bursts, there is a general consensus that 
a highly relativistic flow with large Lorentz factor
$\Gamma\gesim100$ is present. 
Nevertheless, it is difficult to constrain the physics of
the outflow with the available data. Several break frequencies are predicted,
and have been identified in the observations 
\citep[for a review see][]{piran05}.
But direct observation of a power-law spectrum over a substantial 
frequency range is elusive. The \lq\lq measured\rq\rq\ indices frequently
represent interpolations between observation bands. Light curves
are in some cases 
quite accurately known, but can only be used to constrain theories of the
acceleration mechanism when combined with additional assumptions about the
dynamics of the expansion. Nevertheless, all models currently discussed
conform to the trend of a hard electron spectrum at low energy (usually
manifested as a low energy cut-off) and a softer one at high energy, although
in some bursts,
even the X-ray synchrotron emission may lie in the \lq\lq low
energy\rq\rq\ regime according to this definition
\citep{moranetal05}.

A third class of object thought to harbour a relativistic flow 
is that of blazars. 
Relativistic bulk speeds are required both to explain  
apparently superluminal motion 
in their jets and by the rapid variability in their
gamma-ray spectra. As in the case of gamma-ray bursts, the spectra 
do not show clean 
power-law behaviour over several decades of photon energy 
but demand a detailed analysis 
before they can be interpreted in terms of acceleration theory
\citep{sambrunamaraschiurry96,tavecchioetal02}. At
radio frequencies, the spectra are systematically harder than at
high frequencies. This could either 
reflect the intrinsic particle distribution
\citep{mastichiadiskirk97} or be caused by 
internal absorption. Several individual multi-frequency 
flares have been successfully modelled
by using either a very hard injection spectrum 
$s\approx3.3$ combined with a high frequency cut-off  
\citep{konopelkoetal03} or a low frequency cut-off and 
softer $s\approx4.2$ high-energy power-law
\citep{krawczynskicoppiaharonian02}.

The study of Blazar spectra illustrates the difficulties involved in trying to 
extract the intrinsic injected particle spectrum from an observed photon
spectrum that has been shaped by various loss processes and by internal and
possibly external absorption. An interesting alternative 
approach to this problem has
recently been adopted for FR~I radio galaxies by \citet{youngetal05}, who
identify $s=4.1$ as a characteristic power-law index of the acceleration
process. This analysis provides no evidence that the low energy particle
distribution hardens, but cannot be interpreted as
evidence against such a trend.

\section{RELATIVISTIC SHOCKS}
The first-order Fermi mechanism of particle acceleration at relativistic 
shocks rests on the assumption that energetic, charged particles are 
transported stochastically through the background plasma around a shock front
\citep[for a review see][]{kirkduffy99}. Particles cannot be assumed to  
\lq\lq diffuse\rq\rq\ in space. This is because 
the diffusion equation can be derived only 
for particles whose 
angular distribution function is approximately isotropic in the local 
plasma rest frame --- an impossible requirement
if the relative speed of the upstream and downstream plasmas is
comparable to the speed of the particles themselves \citep{kirkschneider87},
as, for example, at a relativistic shock.

\begin{figure}
\begin{center}
\includegraphics[width=6.5 cm]{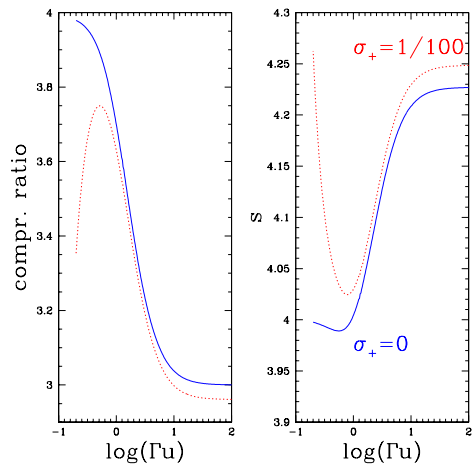}
\end{center}
\caption{\label{indices}%
The power-law index $s$ and the compression ratio of relativistic shocks as
a function of the spatial component of the four-velocity of the upstream
plasma into the shock $\Gamma u$. Two cases are shown: (i) non-magnetised 
with the full Synge/J\"uttner equation of state (solid line) and 
(ii) unmagnetised upstream (dotted line)
but with an oscillating field component generated
to the level of $\sigma_+=1/100$ (see text).
}
\end{figure}
 
Nevertheless, provided the particle transport process is stochastic and
does not introduce a 
characteristic momentum scale into the problem, one can still expect 
the acceleration process to produce a power-law spectrum in particle 
energy, at least for ultra-relativistic particles, whose 
velocity can be considered to be independent of energy. 
A model of the transport process has to be adopted in order to 
find the power-law index, but the result appears 
to be rather insensitive to the particular choice. 
The kinematic problem of particle acceleration at a relativistic
shock, i.e., that of finding the distribution of a collection of test
particles undergoing small-angle, random, elastic (in the plasma frame) 
deflections in the vicinity of a discontinuity in 
the (relativistic) plasma velocity is well-understood.
An analytic method based on an eigenfunction decomposition 
is available which gives the spectrum and angular
dependence of the distribution function at energies well above those
of injection for arbitrary shock speeds \citep{kirketal00}. In
addition, Monte-Carlo
simulations have been performed \citep{bednarzostrowski98,achterbergetal01}
finding good agreement
with the analytic results.  
These are illustrated in Fig.~\ref{indices} 
which shows the compression ratio and the 
high-energy power-law index $s$ as a function of the spatial
component of the 4-speed $\Gamma u$ of the upstream plasma, 
where $\Gamma=(1-u^2)^{-1/2}$.
An interesting aspect of these results is that
the power-law index tends asymptotically 
to the value $s\approx4.23$ for large shock
Lorentz factors (or, equivalently, upstream Lorentz factors), 
independent 
of the equation of state of the plasma. This asymptotic value 
is essentially fixed by the compression ratio of the shock and 
depends only 
weakly on the form of the scattering operator 
used to describe the small-angle deflections.

The eigenfunction expansion method enables the full angle dependence 
of the distribution to be extracted, giving additional insight into the
kinematics of the acceleration process. 
Both upstream and 
at the shock front itself the 
angular dependence is well-approximated by the simple expression
\begin{eqnarray}
f&\propto& \left(1-\mu_{\rm s}u\right)^{-s}
\exp\left(- {1+\mu_{\rm s}\over 1-\mu_{\rm s} u}\right)
\label{eigenfunction}
\end{eqnarray}
where $\mu_{\rm s}$ is the cosine of the angle between the shock normal
and the particle velocity, measured in the frame in which
the shock is at rest and the upstream plasma flows along the shock
normal
at speed $c\vec{u}$. 

\begin{figure}
\begin{center}
\includegraphics[width=6.5 cm]{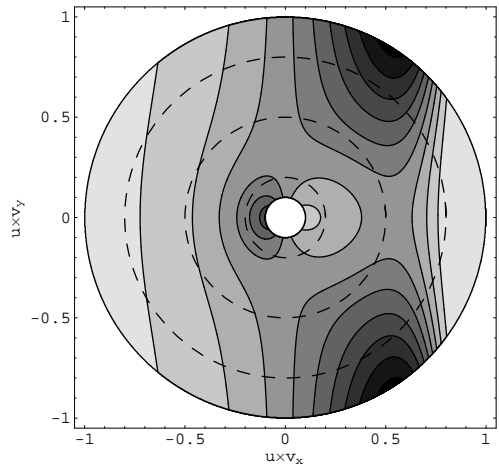}
\end{center}
\caption{\label{polarplot}%
Contour plot of the approximate angular dependence of accelerated particles 
upstream of the shock front (Eq.~\ref{eigenfunction} for $s=4.23$), 
plotted in the rest frame of the shock. 
$v_x$ and $v_y$ specify the particle velocity, and $u$ is the 
three-speed of the 
background 
plasma flow, (which is 
in the positive $x$ direction in this frame) in units of $c$.
Since $v_x^2+v_y^2=1$ for ultra-relativistic particles, the 
angular distribution for given $u$ lies on a circle of radius $u$
(shown as dashed lines for $u=0.2$, $0.5$ and $0.8$). 
The contours are linear,
starting at $f=0.1$ (light) and spaced by 0.1 up to $f=.8$ (dark). 
}
\end{figure}
This function is illustrated in Fig.~\ref{polarplot}
for $s=4.23$. 
Contours of $f$ are shown as a function of a combination of the 
particle velocity components $v_x$ and $v_y$ and
the speed of the upstream plasma into 
the shock $u$. The shock normal lies along the $x$ axis, along which 
the plasma flows. 
Particles with $v_x>0$ are streaming back towards 
the shock front; the distribution is rotationally symmetric about 
the $x$ axis. 
The further one goes away from the shock into the 
upstream region
the better $f$ approximates the full distribution. 
For ultra-relativistic shocks, $f$ is a 
good approximation even at the shock front itself, in which case those 
particles with $v_x>0$ are crossing from upstream to downstream and 
those with $v_x<0$ vice versa.  
For a given upstream 
plasma speed $u$, 
the angular distribution of the accelerated particles is 
determined by the contours intersecting the circle of radius $u$, 
centred on $v_x=v_y=0$. 
At low shock speeds, the distribution function is more or less isotropic, 
with a slight forwards/backwards asymmetry. However, above roughly $u=0.5$, a 
pronounced cone emerges, directed back towards the shock front. 
In the ultra-relativistic limit, this cone has an opening (half)
angle $\theta_{\rm c}$ given by
\eqb
\theta_{\rm c}&=&\arccos\left({s-2\over s}\right)
\eqe
It is independent of the details of the upstream transport, 
provided these can be described as a diffusion in angle. The details of the 
downstream transport enter only in that they have a slight influence on 
the power-law index $s$. 
The cone arises because of two competing physical effects. Firstly, 
those particles that cross the shock 
with velocity inside the cone: $v_x>\cos\theta_{\rm c}$
are depleted because they have a high probability of escape downstream. 
Secondly, the angle at which a particle propagates in the upstream plasma 
reflects the energy and, to some degree, also the angle at which it last 
emerged from the downstream plasma. Figure~\ref{polarplot} shows the upstream 
distribution at fixed energy in the shock rest frame. In this case, the 
higher $v_x$, the greater the energy gain has been since the last crossing. 
A large energy gain also implies that the potential source particles 
are more numerous. Consequently, this effect leads to a steady increase in the 
distribution function as $v_x$ increases.

Using a Monte-Carlo approach, it is possible to
investigate more general forms of the scattering operator, whilst
retaining the effect of a non-vanishing average magnetic 
field \citep{ostrowski93,achterbergetal01,virtanenvainio05}. Provided the
turbulence remains strong, little difference is found. However, as
expected, the acceleration mechanism becomes less effective as the
turbulence diminishes \citep{ostrowskibednarz02}, because the regular 
component of the field in the downstream
region quickly drags particles away from the shock front
\citep{begelmankirk90}.  
Explicit calculations
of particle motion in a 
completely random magnetic field (with vanishing average component)
have been performed by
\citet{ballardheavens92} and \citet{casseetal02}. They have been
used to compute the acceleration around a
relativistic shock for Lorentz factors $\Gamma\le5$
\citep{ballardheavens92} and, more recently, for $\Gamma\le100$
\citep{lemoinepelletier03}. The latter find good agreement with
the analytic result on the asymptotic power-law index. 

Although particle transport in astrophysical plasmas is usually dominated by 
interaction with fluctuations in the electromagnetic
field produced collectively by the background plasma, there
there are strong indications that two-body 
{\em collisional} processes (including those with the photon gas e.g.,
photo-pion production and Compton scattering)  
may be important for the acceleration
and/or the thermalisation of energetic particles in the
inner parts of a GRB fireball $r<10^{16}\,\textrm{cm}$
\citep{derishevetal03,stern03}. These generically 
produce much harder spectra, principally because 
an energetic particle occasionally takes on 
an uncharged \lq\lq identity\rq\rq\ as a photon or neutron. This 
facilitates flights deep into the upstream region, enabling the particle to
profit from an energy boost of a factor $\Gamma^2$, 
which is normally available
to charged particles only on their initial shock encounter. 
However, a hard spectrum
enhances the modification of the 
shock front by the accelerated particles. This implies that 
fully nonlinear calculations will be required to assess the  
importance of collisional processes. 

On the other hand,  
the nonlinear modification of 
a collisionless relativistic shock does
not affect the {\em asymptotic} power-law index. There are two reasons for
this: firstly, isotropised, accelerated particles behave like a
relativistic gas with adiabatic index $4/3$, so that the overall
compression ratio of an ultra-relativistic shock front remains 3,
even when a significant part of the overall energy and momentum flux is
carried by these particles. Secondly, the asymptotic power-law index
in the test-particle picture is {\em soft} (i.e., $s>4$). This means
that it is possible to consider a Lorentz factor above which the
test-particle approximation is valid, because the 
energy density in the remaining accelerated particles is indeed
small. Nevertheless,
a strong nonlinear effect can be exerted by particles of lower
energies, whose mean free path to scattering is comparable to the size of
internal structures in the shock transition \citep{ellisondouble02}.

In order to understand how a 
collisionless, relativistic shock can form, it is necessary
to identify a suitable instability which can lead to dissipation in
the nonlinear regime. 
Currently, the most promising approach to this problem
considers the nonlinear development of the Weibel instability
\citep{yangetal93,yangetal94,medvedevloeb99}, which generates 
downstream magnetic field perpendicular to the streaming motion of the
plasma i.e., in the plane of the incipient shock. 
Particle-in-cell simulations of this situation 
have been performed
\citep{silvaetal03,nishikawaetal03,jaroschekleschtreumann05}
suggesting that magnetic field can be
generated with a strength $\sigma_+$ of a few percent.
(The 
magnetisation parameter $\sigma_+$ is defined as the ratio of the magnetic
energy density to twice the total enthalpy density (including rest
mass) as measured in the downstream plasma rest frame). 
This is encouraging, since it is roughly the level implied by 
spectral modelling of GRB after-glows \citep{panaitescukumar02}. 
However, it has so far not been possible to identify particles that partake in
the first order Fermi process
\citep{hededaletal04}, nor is the ultimate fate of the generated field 
fully understood \citep{medvedevetal05}.

The manner in which magnetic field is generated at the shock
has a strong influence on the spectrum of accelerated particles.
However, if we are
interested only in high energy particles of long mean free path, 
the complex aspects of the problem can be by-passed:
the power-law index predicted by the first-order Fermi mechanism can be
calculated simply by modifying the shock jump conditions to account
for the generated field. To do this, consider time-averaged 
conditions, so that linear functions of the oscillating electromagnetic field
vanish. The stress-energy tensor in the plasma frame is 
\begin{eqnarray}
T^{\mu\nu}&=&\left(w+{B^2\over 4\pi}\right)u^\mu u^\nu +
\left(p+{B^2\over8\pi}\right) g^{\mu\nu}
\nonumber\\
&&- {B^\mu B^\nu\over 4\pi}
\end{eqnarray}
(for notation see \citet{kirkduffy99}) and the last term on the right
hand side does not contribute to the fluxes across the shock front
if the magnetic field lies in the shock plane. As a result, the
jump conditions are the same as those of an unmagnetised fluid,
provided the magnetic enthalpy density $B^2/4\pi$ and pressure $B^2/8\pi$ are
taken into account \citep{lyubarsky03b}. For a relativistic gas, this gives an effective
adiabatic index 
\begin{eqnarray}
\gamma_{\rm eff}&=&{4(1+\sigma_+)\over(3+\sigma_+)}
\end{eqnarray}
leading to an asymptotic compression ratio of 
$1/\left(\gamma_{\rm eff}-1\right)$ and
a relative speed of the upstream medium with respect to the
downstream medium corresponding to the Lorentz factor
$\Gamma_{\rm rel}=\Gamma\sqrt{(2-\gamma_{\rm
 eff})/\gamma_{\rm eff}}$.
As $\sigma_+$ increases, the compression ratio of the shock decreases
 and the high-energy power-law softens, as shown in
 Fig.~\ref{indices}. 
If magnetic field amplification indeed saturates at $\sigma_+\sim1\%$, 
the asymptotic spectral index still remains close to $4.2$.

\section{RELATIVISTIC CURRENT SHEETS}
The first order Fermi process at a collisionless, relativistic shock does 
not appear to produce spectra with $s\lesim4$, so that an additional mechanism
is required in many sources. Since this mechanism tends to manifest itself at
lower energies, it could also play the role of injecting particles into the 
Fermi~I process. 
Possible candidate processes include acceleration by a velocity shear
\citep{stawarzostrowski02,riegerduffy04}, the maser mechanism of 
\citet{hoshinoetal92} and the {\em destruction} of magnetic flux in
the shock front \citep{lyubarsky03b}, as well as the second order Fermi
process of acceleration by a turbulent wave spectrum
\citep{virtanenvainio05}.
Another possibility is acceleration at relativistic
current sheets \citep{kirk04}. As well as its obvious importance 
in the process of flux destruction, this 
possibility is particularly attractive in view of
the fact that field reversals of short length scale 
can be generated at relativistic
shocks \citep{medvedevetal05}.

The current sheets at which reconnection and 
particle
acceleration takes place in astrophysics are relativistic in two senses:
Firstly, the magnetisation parameter,
$\sigma$ is large and the Alfv\'en speed
$v_{\rm A}=c\sqrt{\sigma/(1+\sigma)}$ is close to $c$. Secondly,
the geometry of the current sheet at which magnetic energy is
dissipated and, hence, the field configuration, 
is dictated by a highly relativistic plasma flow.
Particle acceleration depends crucially on both 
the magnetisation parameter and
the field configuration.

The relativistic effects associated with a large magnetisation parameter
are readily appreciated. On the other hand, 
the geometrical effects of a relativistic flow are more subtle.
The situation is closely analogous to that of MHD shock
fronts, which can be classified into \lq\lq subluminal\rq\rq\ 
and \lq\lq superluminal\rq\rq\ according to whether the speed of the
intersection point of the magnetic field and the 
shock front is less or greater than $c$
\citep{drury83,begelmankirk90}. 
In each case, a Lorentz
transformation enables the shock to be viewed from a reference frame
in which it has a particularly simple
configuration: either a de~Hofmann-Teller frame with zero electric
field, or a frame in which the magnetic field is exactly perpendicular
to the shock normal. In the case of a current sheet, the speed of the
intersection point of the magnetic field lines and the sheet
centre-line is important. If it is subluminal, a transformation to a 
de~Hofmann-Teller frame is possible, leading to the 
standard configuration for a nonrelativistic sheet
\citep{chen92,buechnerzelenyi89}. Alternatively, for superluminal motion
of the intersection point, which should be the rule for sheets in
relativistic flows, a frame can be found in which the sheet is a
true neutral sheet with no field lines linking through it. This is, in
fact, the original configuration considered by \citet{speiser65}. For
a relativistic sheet, however, it is the {\em generic} 
case, rather than a very special singular one.

Most discussions of
reconnection treat a Sweet-Parker or Petschek configuration in which 
the length of the current sheet in the average 
field direction determines
the dissipation rate. This is also true for recent analytic treatments
that are relativistic in the sense that the effects of large $\sigma$
are included \citep{lyutikovuzdensky02,lyutikov03,lyubarsky05}. 
But the vanishing of $B_z$ in the generic relativistic case
has important implications, since it is the linking 
field that can eject particles from the sheet, making it 
crucial for the determination of the spectrum
of accelerated particles, 
and, especially, the
maximum permitted energy
\citep{litvinenko99,larrabeelovelaceromanova03}.

\begin{figure}
\begin{center}
\caption{\label{stripedsheet}
The striped pattern of a pulsar wind. A magnetic dipole embedded in
the star at an oblique angle to the rotation axis introduces field
lines of both polarities into the equatorial plane. The current sheet 
separating these regions is shown. In the inset, 
an almost planar portion of this sheet (dashed line) is shown,
together with the magnetic  
field lines, assuming they undergo reconnection.}
\includegraphics[bb=0 0 759 560,width=8.5 cm]{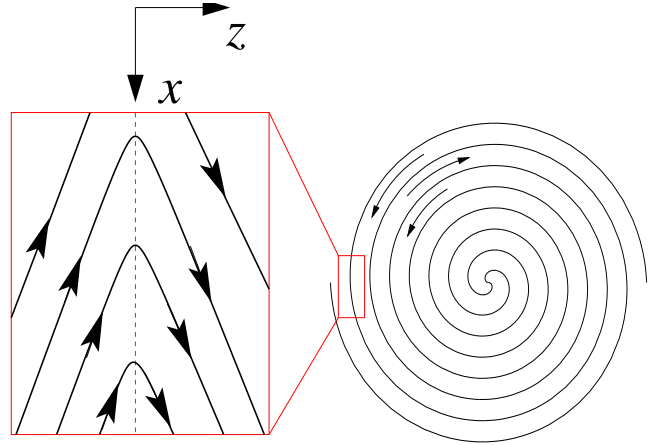}
\end{center}
\end{figure}

Relativistic current sheets, 
can extend over large distances along the field, 
depending on the nature of the boundary conditions. An example, 
drawn from the case of a striped pulsar wind 
\citep{coroniti90,lyubarskykirk01,kirkskjaeraasen03}, is
shown in Fig.~\ref{stripedsheet}. If we assume that reconnection leads
on average to a stationary field configuration, then as the spiral
pattern moves outwards, the linking field lines shown in the inset
must move through the plasma at a speed sufficient to keep their
average distance from the star constant. The striped spiral 
pattern depicted in the figure is
expected to be established well outside the light cylinder, defined to
be at radius $r=r_{\rm L}$, 
where the corotation speed reaches $c$. In this case,
the magnetic chevrons, which must move a distance $2\pi r$ in each
rotation of the spiral pattern, have a superluminal speed equal to
$c r/r_{\rm L}$. Transformation to the frame in which the sheet is a
true neutral sheet involves a small boost in the $x$ direction, and
the resulting configuration has a typical dimension in the azimuthal
direction of $\sim 2\pi r$.

Particle acceleration in current sheets with finite linking field 
($B_z$) has been
extensively investigated \citep{syrovatskii81}. But in the generic,
relativistic, configuration, 
the linking field can play no role in ejecting particles from the sheet. 
Instead,
acceleration is controlled by the finite extent of the sheet 
in latitude, i.e., in the direction parallel to the 
electric field ($E_y$). This is limited 
not by the boundary conditions, but by local parameter values,
as first described by \citet{alfven68}.
Assuming the plasma consists of cold electrons and positrons, 
and that $\sigma\gg1$, the
maximum Lorentz factor $\gamma_{\rm max}$ after acceleration is,
\begin{eqnarray}
\gamma_{\rm max}&=&2\sigma,
\label{pairplasmalimit}
\end{eqnarray}
whereas a cold electron-proton plasma gives
\begin{eqnarray}
\gamma_{\rm max}\,\approx\,\sigma\ &&\textrm{ for protons}
\\
\gamma_{\rm max}\,\approx\,\sigma M/m\ &&\textrm{ for electrons},
\label{leschlimit}
\end{eqnarray}
\citep{kirk04}
with $M$ and $m$ the proton and electron masses, respectively.
It is interesting to note that in a plasma in which the magnetic field
and particle {\em rest mass} are in rough equipartition
($\sigma\approx1$), the upper limit given by Eq.~(\ref{leschlimit})
coincides with that quoted by \citet{leschbirk97}. However, this
situation arises only in relativistic plasmas. In the interstellar
medium, for example, $\sigma\approx10^{-9}$ or smaller, in which case 
the upper limit on the energy gain reduces to $M v_{\rm A}^2$. 
Standard estimates of the interstellar magnetic field and particle density 
($1\,\mu$G, $1$~proton/cm$^{3}$) imply that electrons can be
accelerated, at most, to
only mildly relativistic energies.
In this case, and in solar system applications, direct acceleration by
the DC field may be masked by particle acceleration in the turbulence
fed by reconnection or the associated shocks \citep{cargill01}.
 
The picture sketched above applies only to quasi-steady current
sheets. However,
\citet{zelenyikrasnoselskikh79} have shown that 
relativistic current sheets are unstable to the growth of
the tearing mode and other
instabilities are also likely to 
operate (see, for example, \citet{daughton99}). 
On scale lengths comparable to the sheet thickness  
an unsteady, oscillating component of $B_z$ may be generated.
Thus, locally, the nonrelativistic picture may be relevant
to the micro-structure of the sheet, although not in its standard
2-dimensional stationary incarnations. 

Particle-in-cell simulations can provide valuable insight here,
provided they account for relativistic particle motion. Such simulations
been performed in 3D by
\citet{jaroscheketal04} and \citet{zenitanihoshino05}, who noted 
the growth of corrugations in the current sheet in
the direction of the electric field $E_y$ and  
identified them as due to the relativistic 
drift kink instability. \citet{jaroscheketal04} found a 
very hard spectrum of energetic particles, that can be
understood in terms of an  
\lq\lq acceleration zone\rq\rq\ near the sheet centre, 
in which the electric field exceeds the
magnetic field.
The escape rate from this zone is then approximated as
the time taken by a particle to complete one quarter of a revolution
around the linking component of the magnetic field
\citep{zenitanihoshino01}. However, the role of the kink instability appears
to place a relatively modest maximum energy limit on the acceleration process.

\section{Summary}
Although the details of the plasma physics remain obscure, simple
kinematic considerations suggest that acceleration at shocks
imprints a characteristic power-law index on the particle spectrum. 
In the
case of relativistic shocks, 
it is $p^{-4.2}$, and seems not to be sensitive to nonlinear 
effects, or the effects of magnetic field generation at the level 
of $\sigma_+\sim\%$. 

A spectrum consistent with this prediction 
has been identified in a few objects, but 
observations also show that acceleration into a much harder spectrum
is needed at low energies. Current sheets are in principle capable of 
producing particles with such a spectrum, but 
a full understanding of the way they operate remains a challenging goal.
%

%
%




\end{document}